\documentclass[article]{WileyASNA-v1}

\usepackage{graphics,graphicx}
\usepackage{float}
\usepackage{amsmath}
\usepackage{subfig}
\usepackage{multirow}
\usepackage{amssymb}
\usepackage{dirtytalk}
\articletype{Proceeding}%

\received{18 October 2021}
\revised{18 October 2021}
\accepted{18 October 2021}

\begin{document}


\title{The central FR0 in the sloshing cluster Abell 795 - indications of mechanical feedback from \textit{Chandra} data}

\author[1,2]{F. Ubertosi*}

\author[1,3]{M. Gitti}

\author[2]{E. Torresi}

\author[1,4]{F. Brighenti}

\author[2]{P. Grandi}

\authormark{F. UBERTOSI \textsc{et al}}

\address[1]{Dipartimento di Fisica e Astronomia (DIFA), University of Bologna, via Gobetti 93/2, I-40129 Bologna, Italy}

\address[2]{Istituto Nazionale di Astrofisica (INAF) - Osservatorio di Astrofisica e Scienza dello Spazio (OAS), via Gobetti 101, I-40129 Bologna, Italy}

\address[3]{Istituto Nazionale di Astrofisica (INAF) - Istituto di Radioastronomia (IRA), via Gobetti 101, I-40129 Bologna, Italy}

\address[4]{University of California Observatories/Lick Observatory, Department of Astronomy and Astrophysics, University of California, Santa Cruz, CA 95064, USA}

\corres{*Francesco Ubertosi. \email{francesco.ubertosi2@unibo.it}}


\abstract{We present a detailed study of the galaxy cluster Abell 795 and of its central Fanaroff-Riley Type 0 (FR0) radio galaxy. From an archival \textit{Chandra} observation we found a dynamically disturbed environment with evidences for sloshing of the intracluster medium. We argue that the environment alone cannot explain the compactness of the radio galaxy, as similar conditions are also found around extended sources. We identified a pair of putative X-ray cavities in the proximity of the center: these could have been created in a past outburst of the FR0, and dragged away by the large scale gas movement. The presence of X-ray cavities associated with a FR0 could open a new window on the study of jet power and feedback properties of this recently discovered class of compact radio galaxies.}


\jnlcitation{\cname{%
		\author{Ubertosi F.}, 
		\author{Gitti M.}, 
		\author{Torresi E.}, 
		\author{Brighenti F.}, and 
		\author{Grandi P.}} (\cyear{2021}), 
	\ctitle{The central FR0 in the sloshing cluster Abell 795 - indications of mechanical feedback from \textit{Chandra} data}, \cjournal{Astron. Nachr}, \cvol{1}. https://doi.org/10.1002/asna.20210055}


\maketitle


\section{Introduction}\label{sec1}
In the most relaxed galaxy clusters, the intracluster medium (ICM) cools and accumulates to the center, building high density gas reservoirs in the proximity of the brightest cluster galaxy (BCG). The cold central gas is thought to fuel the supermassive black hole (SMBH) in the BCG, which in turn launches powerful jets capable of excavating depressions (the so-called X-ray cavities) in the ICM ~\citep{2007ARA&A..45..117M,2012NJPh...14e5023M}. The tight interplay between the ICM thermodynamical state and the activation of the central active galactic nucleus (AGN) points to the existence of a \textit{feedback loop}, which has been validated with multi-wavelength observations (e.g., ~\citealt{2001MNRAS.328..762E,2004ApJ...607..800B,2006PhR...427....1P}). \\ However, the details of the mechanism are currently unknown. Indeed, in relaxed clusters it is possible to observe spirals, ripples, and discontinuities in surface brightness (e.g., \citealt{2010A&A...516A..32G}): these features are thought to be caused by \textit{sloshing}, i.e. an oscillation of the cold gas in the potential of the cluster following a minor merger, which results in the formation of \textit{cold fronts} (surface brightness discontinuities, with the inner side being colder and denser than the outside; see e.g., for reviews \citealt{2007PhR...443....1M,2016JPlPh..82c5301Z}). Considering the link between ICM cooling and AGN activity in clusters, the displacement of the gas from the BCG could impact the stability of the cooling cycle, and the sloshing-induced turbulence might interfere with the jet propagation. The outcomes of sloshing motion on AGN fuelling and jet stability have been investigated \citep{2019ApJ...885..111P,2020MNRAS.496.1471K} in the context of extended radio galaxies, but not for smaller ones. \\ While BCGs typically host Fanaroff-Riley Type I AGNs (FRIs, e.g., \citealt{2009ApJ...704.1586S}), the observation of radio galaxies in the mJy regime (e.g., \citealt{2012MNRAS.421.1569B}) has unveiled that the radio loud AGN population at z$<$0.05 is dominated by low-luminosity radio compact objects, being unresolved at the 5$''$ resolution of the FIRST (Faint Images of the Radio Sky at Twenty centimeters survey, \citealt{1995ApJ...450..559B}.) survey. Among these, \citet{2015A&A...576A..38B} selected AGNs associated with red massive early type galaxies, with a high mass SMBH ($\ge$$10^{8} $M$_{\odot}$), spectroscopically classified as low excitation galaxies; being apparently different from FRIs only in radio size, these sources have been named Fanaroff-Riley Type 0 (FR0) radio galaxies \citep{2011AIPC.1381..180G}. High-resolution radio observations have confirmed the compact morphology of FR0s (e.g., \citealt{2018ApJ...863..155C}), which is maintained also at lower frequencies \citep{2019A&A...631A.176C}, thus excluding that these radio galaxies are fading, once extended FRIs.\\ It has been proposed that FR0 could be powered by slowly spinning SMBH launching unstable jets (e.g., \citealt{2015A&A...576A..38B}). \cite{2020A&A...633A.161C} found that FR0s tend to reside in clusters and groups, but with a smaller number of members on average than FRIs: the authors proposed that this difference could lower the probability of SMBH mergers, and prevent the AGN from spinning up its engine. The presence of FR0s in galaxy clusters and groups opens the possibility of studying their environments in the X-ray band, in order to perform comparisons with the cluster environment of extended radio galaxies. Additionally, it could be possible to obtain information on the jet power of FR0s by studying the interaction between the radio galaxy and the surrounding ICM. \\
\textbf{Our target}. To investigate this subject, we performed a detailed analysis of the FR0 in Abell 795, a galaxy cluster at a redshift z$\sim$0.137 with a 30 ks archival \textit{Chandra} observation (ObsID 11734, see \url{https://cda.harvard.edu}). The FR0 coincides with the BCG of A795 \citep{2018MNRAS.476.5535T}, located at RA, DEC: 09:24:05.3, +14:10:21.5 (J2000). Since the central regions of clusters are distinctive in terms of ICM density and dynamics, the \textit{Chandra} observation of A795 offers the possibility of studying the link between ICM cooling and AGN activity for a compact radio galaxy. The FR0, hosted in a passive elliptical galaxy, is powered by a 3.9$\times10^{8}$M$_{\odot}$ SMBH with an accretion efficiency  of $\sim$10$^{-3}$ \citep{Ubertosi2021}, typical of low excitation radio galaxies. The compactness of the AGN is evident from radio observations: the radio galaxy is unresolved by FIRST at 1.4 GHz, implying an upper limit on the size of $\sim$10 kpc. The core of the radio galaxy has been barely resolved with the VLA at 8.4 GHz (0.2$''$ resolution, \citealt{2015MNRAS.453.1201H}). A MERLIN observation at 5 GHz (sub-arcsec resolution, \citealt{2010MNRAS.408.2261K}) revealed a possible core-jet morphology, with the putative jet oriented towards noth-east (position angle $\sim$120$^{\circ}$), and with a largest linear size of $\sim$1 kpc. 
\\ In this work we adopt the following cosmology: $H_{0}$ = 70 km s$^{-1}$ Mpc$^{-1}$, $\Omega_{\text{m}}$=0.3, $\Omega_{\Lambda}$=0.7, which results in a conversion of 2.43 kpc/$''$. Uncertainties are reported at the $1\sigma$ confidence level. The radio spectral index $\alpha$ is defined as $S_{\nu}\propto\nu^{-\alpha}$ (where S is the flux density and $\nu$ is the frequency). The \textit{Chandra} data reduction and spectral fitting of A795's observation were carried out using \texttt{CIAO-4.12} and \texttt{Xspec-v12.10}, respectively (see \citealt{Ubertosi2021} for details).

\section{Analysis of the ICM}\label{sec2}
\begin{figure*}[h]
	\centering
	\includegraphics[width=1\hsize]{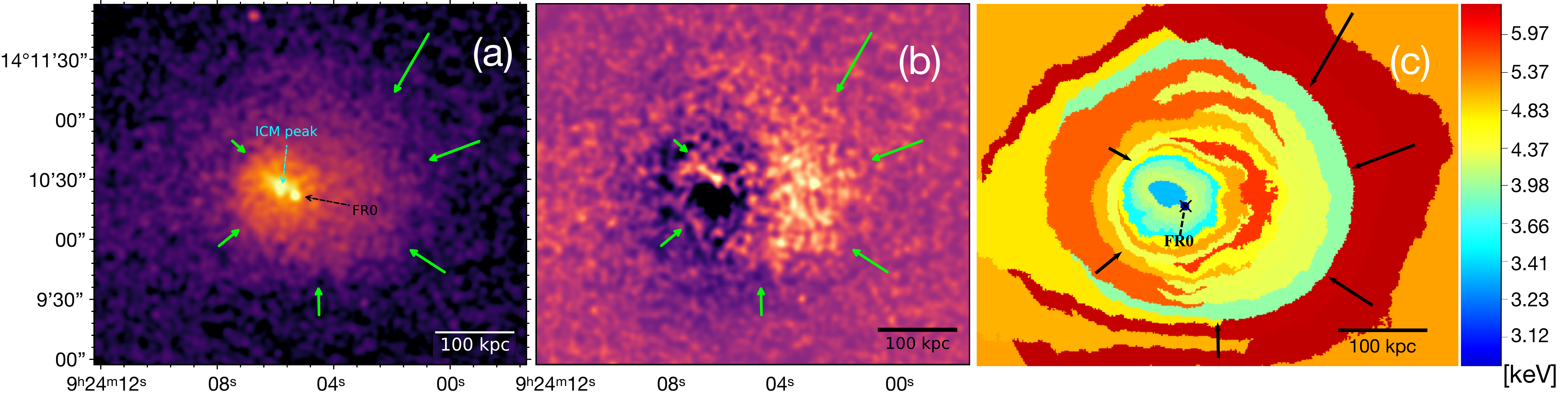} 
	\caption{\textit{Panel a:} \textit{Chandra} image (0.5-2 keV) of A795, Gaussian-smoothed with kernel radius of 3$''$. The ICM peak and the position of the FR0 in the BCG are indicated. \textit{Panel b:} 2D $\beta$ model subtracted \textit{Chandra} image over the same region of panel \textit{a}. \textit{Panel c}: Temperature map of A795; relative errors on temperature are of $\approx$25\%. In each panel, the arrows (green in \textit{a} and \textit{b}, black in \textit{c}) highlight the spiral geometry (see Sect.\ref{sec2}).}
	\label{fig:A795wide}
\end{figure*}

The 0.5-2 keV \textit{Chandra} image of A795 (Fig.\ref{fig:A795wide}, \textit{a}) reveals the presence of surface brightness discontinuities, to the east and west of the core. In order to enhance these edges, we fitted the image with a 2D $\beta$-model, and then subtracted the model from the image. The resulting residual map of Fig.\ref{fig:A795wide}\textit{b} shows that the ICM is arranged in a spiral geometry which extends to $\sim$200 kpc from the center. The spiral is particularly enhanced in two opposite arcs, which coincide with the edges seen in the original image. 
\\ Fig.\ref{fig:A795wide}, \textit{c} shows a temperature map of A795, obtained by binning the \textit{Chandra} image with the \texttt{CONTBIN} algorithm \citep{2006MNRAS.371..829S} in order to reach a signal-to-noise ratio of 30, and fitting the ICM spectrum of each region with an absorbed thermal model (\textbf{tbabs}$\ast$\textbf{apec}). The temperature map shows that the coldest gas nicely follows the spiral shape of the ICM, suggesting that the two edges could be cold fronts.
To measure the thermodynamical properties of each jump, we extracted spectra from three sectors: the first following the discontinuity, the second enclosing the region outside the edge, and the third extended to the edge of the chip (to account for the ICM along the line of sight). The resulting spectra were fitted with a deprojected, absorbed thermal model (\textbf{projct}$\ast$\textbf{tbabs}$\ast$\textbf{apec}). We found that the inner side of each front has a lower temperature and a higher density than the outer side:
\begin{enumerate}
	\item  A temperature gradient $T^{\text{out}}/T^{\text{in}} $=2.07$\pm$0.53 and a density jump $n_{e}^{\text{in}}/n_{e}^{\text{out}}$=2.67$\pm$0.10 for the East Front ($\chi^{2}$/D.o.f.=344.8/388).
	\item A temperature gradient $T^{\text{out}}/T^{\text{in}}$=1.61$\pm$0.46  and a density jump $n_{e}^{\text{in}}/n_{e}^{\text{out}}$=2.66$\pm$0.13 for the West Front ($\chi^{2}$/D.o.f.=364.1/377).
\end{enumerate}
These results confirm the cold front nature of the discontinuities. Therefore, our analysis unveiled that A795 is a disturbed cluster, in which the coldest gas phase has been displaced from the center and is now oscillating over the cluster scale. 


\section{The FR0 complex environment}\label{sec3}
\begin{figure}[h]
	\centering
	\includegraphics[width=0.92\hsize]{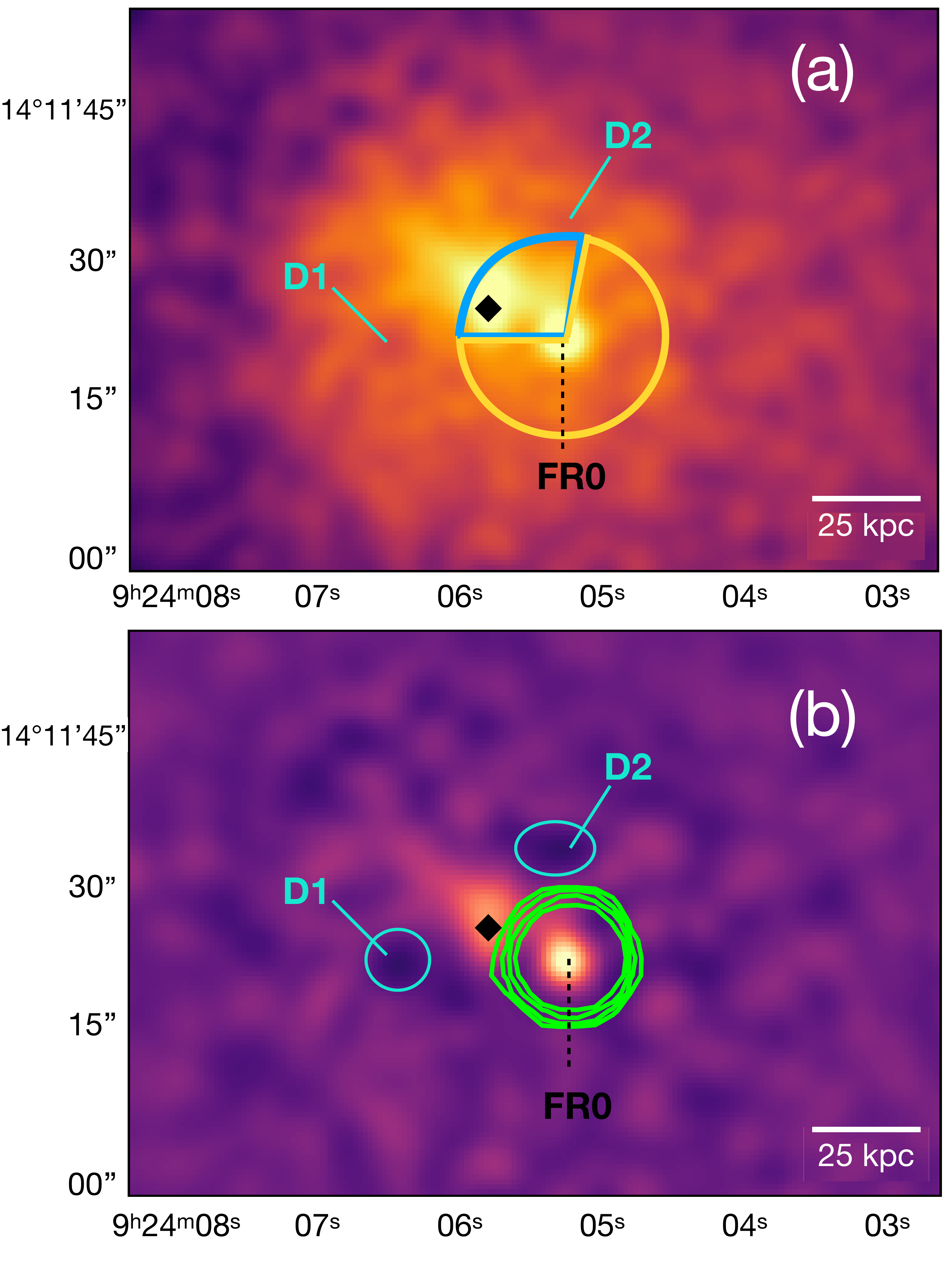} 
	\caption{\textit{Panel a:} Image of the cluster core, with regions used to study the ICM properties around the FR0 overlaid in blue and yellow (see Sect.\ref{sec4}). \textit{Panel b:} 1$''$- 5$''$ unsharp mask image with FIRST radio contours at 1.4 GHz overlaid in green. In both panels the black square marks the ICM peak, while the position of the X-ray cavities is indicated in cyan.}
	\label{fig:A795zoom}
\end{figure}

To investigate the interplay between the cluster environment and the central FR0, we determined the condition of the ICM within $\sim$30 kpc from the BCG. Fig.\ref{fig:A795zoom}, \textit{a}) shows a \textit{Chandra} image of the cluster core, where it is possible to spot the FR0 as a bright point source and the ICM peak on its left. While in relaxed clusters the ICM is centered on the BCG, in A795 there is an offset of $\sim$18 kpc between the two centers. We interpreted this separation as a consequence of the ICM displacement from its original \textit{relaxed} configuration; in particular, this offset suggests that sloshing is present around the BCG. \\The aim of our analysis was to probe whether the ICM density and temperature nearby the FR0 are peculiar w.r.t. those found around extended FRI radio galaxies in clusters, which would suggest that the cluster environment could be playing a major role in determining the size of the radio galaxy.
Considering the geometry of the cold gas spiral, we speculated that the north-east side of the AGN could consists of lower temperature gas w.r.t. the south-west side. Therefore, to investigate the properties of the ambient gas, we extracted the spectrum of the ICM surrounding the FR0 from two annular sectors of inner radius 2'' and outer radius 12'', centered on the BCG (see Fig.\ref{fig:A795zoom}, \textit{a}). Fitting with a \textbf{tbabs}$\ast$\textbf{apec} model returned a temperature kT=3.54$^{+0.41}_{-0.33}$ keV for the NE sector, a temperature kT=4.30$^{+0.30}_{-0.29}$ keV for the SW one, and an average density $n_{e}=2.1\pm0.1\times10^{-2}$ cm$^{-3}$. These results confirm the presence of temperature gradients around the BCG, with the colder phase being spatially connected to the sloshing spiral.
The spectral analysis unveiled that the environment of this FR0 is disturbed, since cold gas movements have shaped the morphology of the cluster. However, it is crucial to note that these conditions are not peculiar: sloshing has been observed in galaxy clusters with extended FRIs at their center, whose jets can be bent by the gas movement (e.g., \citealt{2020MNRAS.496.1471K}) but not destroyed. Moreover, the density of the central ICM is typical of galaxy clusters cores ($n_{e}\approx$10$^{-2}$ cm$^{-3}$, e.g., \citealt{2007ARA&A..45..117M}), thus excluding that the X-ray emitting gas is hampering the jet propagation. Hence, unless the jets of this FR0 are intrinsically weak, sloshing alone cannot explain the small size of the radio galaxy. 

\section{X-ray cavities in the ICM}\label{sec4}
The \textit{Chandra} image of the cluster core (Fig.\ref{fig:A795zoom}) shows hints of a pair of depressions (D1 and D2) on opposite sides of the center. To highlight these features, we produced an unsharp masked image by subtracting a 5$''$ Gaussian smoothed image from a 1$''$ Gaussian smoothed one. The resulting map (Fig.\ref{fig:A795zoom}, \textit{b}) emphasizes the two depressions: the structures are slightly elliptical, with an average radius of $\sim$5 kpc, and lie at $\approx$30 kpc from the FR0 (see Tab.10 from \citealt{Ubertosi2021}). The significance of the depressions (30\% less counts than their surroundings) is at $\sim$2$\sigma$, therefore we classified D1 and D2 as \textit{putative X-ray cavities}. Considering the radio compactness of the FR0, and the lack of information on the duty cycle and feedback properties of these radio galaxies \citep{2019MNRAS.482.2294B,2020A&A...633A.161C}, the discovery of the X-ray cavities is puzzling. With the available radio data on the FR0, there are no indications that the cavities are radio filled. We note that if the plasma inside the depressions has aged, low frequency radio observations could detect it. \citet{Ubertosi2021}, by inspecting low frequency observations of A795, discovered a steep spectrum, diffuse source centered on the cluster; however, the resolution of the available data do not allow to investigate the presence of radio emission coincident with the cavities. Alternatively, the radio surface brightness inside the depressions could lie below the sensitivity of the available observations. Interestingly, the total fluxes measured by the NVSS (National Radio Astronomy Observatory VLA Sky Survey, \citealt{1998AJ....115.1693C}) survey (45$''$ beam) and by the FIRST survey (5$''$ beam) at 1.4 GHz differ by $\sim$5 mJy. This excess flux could arise from low surface brightness extensions coincident with the cavities, but the resolution of the NVSS provides no clue on the excess morphology. On the contrary, core variability might explain this difference.  
\\An additional peculiarity is that the cavities are not on opposite sides of the AGN, but they are offset towards north-east. Since this is also the direction of sloshing motions in the central regions (see Fig.\ref{fig:A795wide}), we speculated that the cavities might have been inflated in the past in the proximity of the FR0 ($\sim$1-10 kpc from the core). Later, the gas turbulent motion could have dragged the cavities away towards north east ($\sim$30 kpc from the core).  

\section{Summary \& Conclusions}\label{sec5}
The \textit{Chandra} observation of A795 revealed the presence of sloshing movement of the ICM around the FR0 in the BCG. However, the overall thermodynamical conditions do not differ from typical FRI-cluster environments, which highlights the role of an intrinsic jet weakness in explaining the size of the FR0. A pair of X-ray cavities was found in the proximity of the FR0, and their position w.r.t. the FR0 is consistent with a sloshing-influenced uprise of the bubbles. If the connection between the cavities and the central AGN is confirmed by future deep X-ray and radio observations, the BCG of A795 would be the first discovered FR0 to have established a feedback cycle in a galaxy cluster. 

\section*{Acknowledgments}
Open Access Funding provided by Universita di Bologna within the CRUI-CARE Agreement. WOA Institution: Universita di Bologna. Blended DEAL: CARE.


\bibliography{proceeding.bib}%

\section*{Author Biography}

\begin{biography}{}{\textbf{Francesco Ubertosi} is a PhD student in Astrophysics at the University of Bologna, under the supervision of Prof. M. Gitti and Prof. F. Brighenti. In 2021 he participated at the 6th Workshop on Compact Steep Spectrum and GHz-peaked spectrum Radio Sources.}
\end{biography}

\end{document}